\documentclass[aps,prl,twocolumn,notitlepage,superscriptaddress,nogroupedaddress]{revtex4-2}
\usepackage{graphicx} 
\usepackage{physics}
\usepackage{hyperref}
\usepackage{amsmath}
\usepackage{xcolor}

\begin{document}

\title{Collective preparation of large quantum registers with high fidelity}
\author{Lorenzo Buffoni}
\affiliation{Department of Physics and Astronomy, University of Florence, 50019 Sesto Fiorentino, Italy}
\author{Michele Campisi}
\affiliation{NEST, Istituto Nanoscienze-CNR and Scuola Normale Superiore, I-56127 Pisa, Italy}

\begin{abstract}
We report on the preparation of a large quantum register of 5612 qubits, with the unprecedented high global fidelity of $F\simeq 0.9956$. This was achieved by applying an improved cooperative quantum information erasure (CQIE) protocol [Buffoni, L. and Campisi, M., Quantum \textbf{7}, 961 (2023)] to a programmable network of superconducting qubits featuring a high connectivity. At variance with the standard method based on the individual reset of each qubit in parallel, here the quantum register is treated as a whole, thus avoiding the well-known orthogonality catastrophe wehereby even an extremely high individual reset fidelity $f$ results in vanishing global fidelities $F=f^N$ with growing number $N$ of qubits.
\end{abstract}

\maketitle
Quantum computers are inherently susceptible to external disturbances that greatly hinder the possibility of achieving quantum advantage.
For this reason, quantum computing research has focused on error correction \cite{bluvstein2024logical,shor1995scheme,google2023suppressing} and mitigation \cite{cai2023quantum}, fault tolerance \cite{aharonov1997fault} and other techniques aimed at achieving higher fidelities in the output state of quantum computations.

Fidelity may be degraded because of various sources acting at different stages in the course of a quantum computation. In this regard, it is useful to distinguish between the sources of error incurred during the execution of quantum circuits, (i.e. gate noise) and SPAM (state preparation and measurement) errors. Here we focus on the latter, and in particular  we concentrate on the problem of preparing a quantum register composed of many qubits in a high fidelity state. 

Efforts in that direction have been aimed so far exclusively at increasing individual qubit preparation fidelity. However, that is not a scalable strategy;
as thoroughly discussed in \cite{buffoni2022third},
if each qubit is individually prepared with a fidelity $f=(1-\varepsilon)$, then the total register fidelity $F$ would dramatically diminish with the number of qubits, $N$:
\begin{align}
 F=f^N=(1-\varepsilon)^N    \, ,
 \label{eq:F=f^N}
\end{align}
a phenomenon analogous to Anderson ``orthogonality catastrophe'' \cite{anderson1967infrared}. For example, with the current state of the art SPAM fidelity of $f=0.997$ for superconducting qubits \cite{opremcak2021high}, the global fidelity would drop to $F \simeq 0.45$ with $N=262$, and $F\simeq 4.7 \times 10^{-8}$ for $N=5612$.

In order to overcome this major block, here we abandon the paradigm of $N$ individual qubit resets occurring in parallel and embrace instead a new paradigm based on collective effects, to achieve a genuinely global preparation of the whole quantum register. The quest for collective advantages is currently in the limelight of intense research in stochastic and quantum thermodynamics \cite{rolandi2023collective}, with applications that span from metrology, \cite{Giovannetti06PRL96},
to quantum batteries \cite{Alicki13PRE87,Hovhannisyan13PRL111,Binder15NJP17,Ferraro18PRL120,campaioli2017enhancing,Andolina19PRL122,Gyhm22PRL128,mukherjee2021many,Qach22SCIADV8},
(quantum) heat engines,
\cite{Campisi16NATCOMM7,Jaramillo16NJP18,Holubec17PRE96,Niedenzu18NJP11,Herpich18PRX8,vroylandt2018collective,Kloc19PRE100,GelbwaserKlimovsky19PRA99,Watanabe20PRL124,Abiuso20PRL124,Latune20NJP8,Niedenzu18NJP11,Fogarty20QST6,Kloc21PRAPP16,Barra22NJP24,Souza22PRE106,Piccitto22NJP24,Filho23PRR5,Kolisnyk23PRAPP19,Solfanelli23NJP25,Jaseem23PRA107}
and quantum transport
\cite{Schaller16PRE94,Kamimura23PRL131,Andolina24arXiv2401.17469}.
While this activity is almost exclusively of theoretical character (Ref. \cite{Qach22SCIADV8} is an exception), the present study is one of the first to practically demonstrate a collaborative advantage on a physical quantum hardware, and the first to demonstrate its application to high fidelity quantum register preparation.

In a previous work \cite{buffoni2023cooperative} we have demonstrated a multi-qubit reset mechanism where an ensemble of $N=256$ qubits starting in a completely mixed state was brought close to a pure state featuring all qubits being in the computational state $|0\rangle$. The method, dubbed ``cooperative quantum information erasure'' (CQIE),
uses both cooperative and quantum effects to achieve the record \emph{erasure action} (i.e., the product of energy cost per qubit and reset time) of 
$10^{-22}$ erg s/bit occurring (within our experimental error) at Landauer cost of $kT \ln 2$ per bit.

Here we demonstrate the employment of CQIE to prepare large quantum registers with unprecedented global fidelity. The CQIE method employed here is a variant of that originally employed in \cite{buffoni2023cooperative}, featuring now a more effective choice of protocol parameters and their time-dependence.
We implemented it on a programmable network of superconducting qubits, specifically, the D-Wave Advantage 6.4 quantum annealing processor \cite{dwdocs} and 
achieved exceptionally high values of the measured global fidelity. 
For example, we demonstrate a global SPAM fidelity of  $F\simeq 0.9996$ on a register of $262$ qubits, and $F \simeq 0.9956$ with as many as $5612$ qubits. To the best of our knowledge such high values of global SPAM fidelity were never reported in the literature for such large qubit registers.

We compared the results obtained for various network sizes (up to $N=5612$) with those obtained from a variant of CQIE featuring non-interacting qubits, that implements a non-cooperative, parallel reset of each qubit. While the global fidelity dramatically decreased in this case, as predicted by Eq. \ref{eq:F=f^N}, it remained almost constant with increasing $N$ in the cooperative case, thus demonstrating the genuinely cooperative nature
of the method. This was corroborated by the observation that CQIE global resets on networks with intermediate levels of connectivity results in intermediate values of global fidelity.

\section{CQIE working principle}
CQIE is a method to reset a network of qubits to a state that is as close as possible to the pure state featuring each and all qubits being in their computational state $|0\rangle$, regardless of its initial state \cite{buffoni2023cooperative}. CQIE leverages cooperative, quantum and dissipative phenomena to achieve that. The idea is that the quantum register is a spin network featuring spontaneous symmetry breaking, subject to an external magnetic field, i.e. a many-body quantum system with Hamiltonian 
\begin{align}
    H = - \mathcal{B}_x \sum_{i=1}^N \sigma_i^x- \mathcal{B}_z \sum_{i=1}^N \sigma_i^z - \mathcal{J} \sum_{i,j} \alpha_{i,j}\sigma_i^z \sigma_j^z \, .
\end{align}
The coefficients $\alpha_{i,j}$ define the adjacency matrix of the network's graph, and the register is immersed in an environment at temperature $T$.
The method consists in changing the three parameters $\mathcal{B}_x,\mathcal{B}_z,\mathcal{J}$ in time in a cyclical fashion, according to the following general scheme \cite{buffoni2023cooperative}. The protocol starts with no magnetic fields $\mathcal{B}_x,\mathcal{B}_z$ at a value of $\mathcal{J}$ above the critical value $\mathcal J_C$ separating the ferromagnetic phase, $\mathcal J>\mathcal J_C$, from the paramagnetic phase, $\mathcal J<\mathcal J_C$ (note that $\mathcal J_C$ generally depends on the environment's temperature $T$). One then decreases $\mathcal{J}$ so as to decrease the height of the Landau free energy barrier thus allowing the system to get demagnetised. This de-magnetisation is highly enhanced by turning on a transverse field $B_x$, that allows tunnelling through the barrier. Once in the paramagnetic phase, one goes back to the initial point, but now adding some longitudinal field $B_z$, to drive the network to a state of high net magnetisation along the direction of $B_z$, regardless of its initial state.

As demonstrated in  \cite{buffoni2023cooperative} the method works excellently. On a 2D Ising network of $N=256$ superconducting qubits with nearest neighbour interaction, we demonstrated that each qubit in the network was reset to the $|0\rangle $ state with an individual fidelity of $f=0.999$, while the global network fidelity $F$ which is the focus of the present work was not addressed.

The cooperative nature of the CQIE method comes from the physics of spontaneous symmetry breaking and is enabled by qubit-qubit interactions. The quantum nature comes from quantum tunnelling and is enabled by the transverse field $\mathcal B_x$. Dissipation also plays a crucial role as it is impossible to increase the purity of the quantum register if it evolves unitarily, i.e., without interacting with an environment: the latter provides in fact the bin where to dump the register mixedness (i.e. its ``impurity'').

\section{Improving the CQIE cycle} 
\begin{figure}
    \centering
    \includegraphics[width=\linewidth]{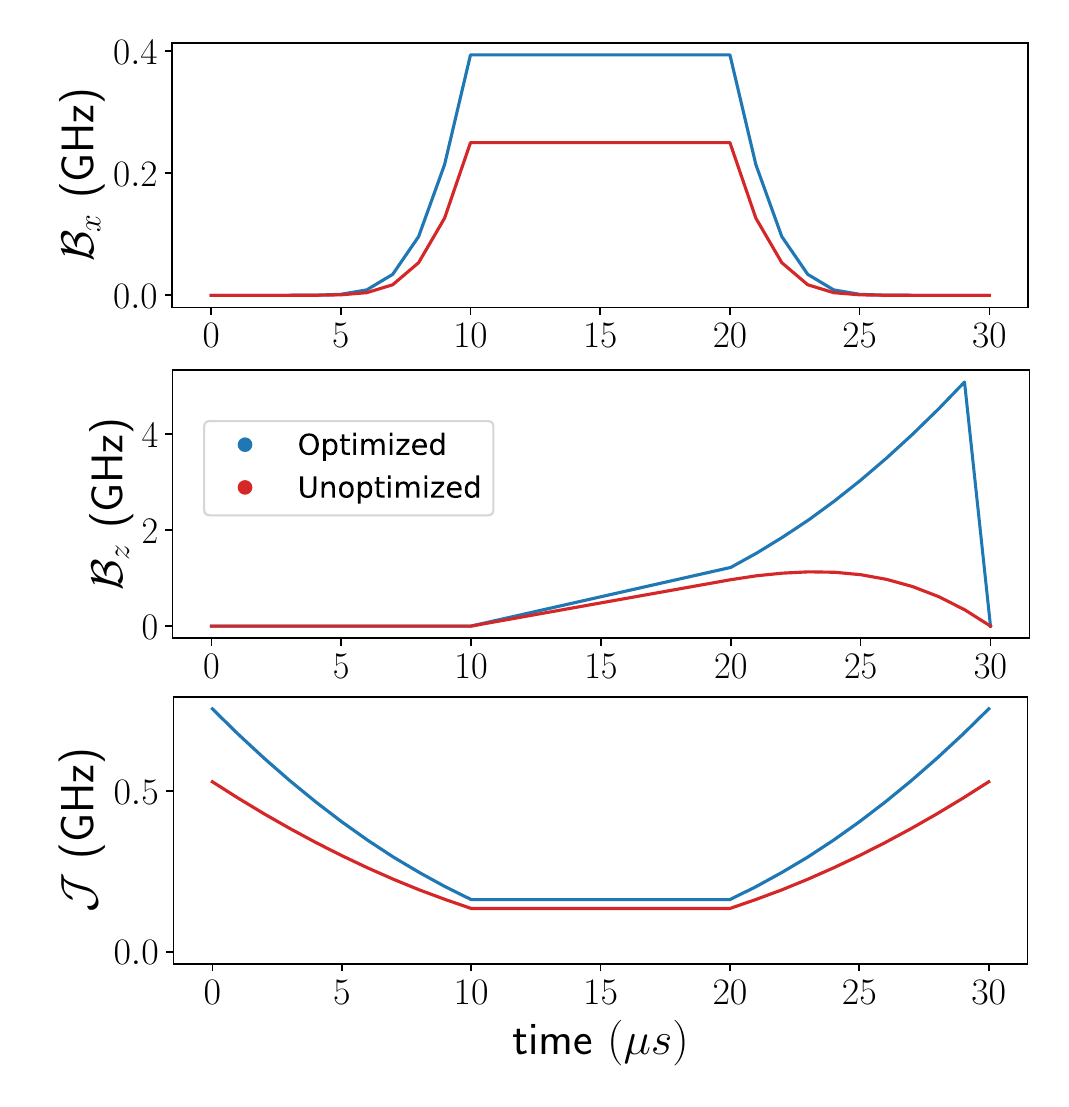}
    \caption{(Red) Original (unoptimized) CQIE protocol \cite{buffoni2023cooperative}. (Blue) Optimized CQIE protocol.
    }
    \label{fig:protocol}
\end{figure}

Figure \ref{fig:protocol} shows the cyclic time-dependent protocol, $\mathcal{B}(t), \mathcal{B}_z(t),\mathcal{J}(t)$, that was employed in Ref. \cite{buffoni2023cooperative} to implement CQIE on a programmable network of superconducting qubits (i.e., a quantum annealer), see the red curves.
As stressed in \cite{buffoni2023cooperative} that cycle is only one of the many cycles that one can employ to implement CQIE. There is in fact plenty of room for optimising the cycle in order to maximise the effectiveness of the CQIE protocol. While a systematic optimisation is beyond the scope of the present study, we nonetheless proceeded to a heuristic improvement of the path, based on trial and error runs

These runs were implemented on a D-Wave Advantage 6.3 processor that implements a connectivity graph with up to $15$ couplings per qubit called Pegasus topology \cite{pegasus}. The variation in time of the control fields can be tuned with a resolution up to $0.01 \mu s$ on these processors.
The complete results of the optimization runs are reported in the appendix. They provided us with three critical insights. The first is that the higher the longitudinal ($\mathcal{B}_z$) field, the higher the final magnetization. Second, the later and the faster that field is turned off the more effective the protocol is. The third is that the transverse field ($\mathcal{B}_x$) is so effective at de-correlating the qubits that it is in fact not necessary that the critical point $\mathcal{J}_C$ be crossed. 

These three key insights led us to define a protocol of duration $\tau=30 \mu s$ with stronger interactions than the original CQIE, a stronger transverse field and a higher longitudinal field. 
This improved protocol, depicted in blue in Fig.\ref{fig:protocol}, was employed to prepare large quantum registers with unprecedented high fidelity, as discussed below.

Note that the colour coding used in Fig.\ref{fig:protocol} is used in all figures of the manuscript: red curves refer to the un-optimised protocol, blue curves refer to optimised protocol.

\section{Results}

\begin{figure}
    \centering
    \includegraphics[width=\linewidth]{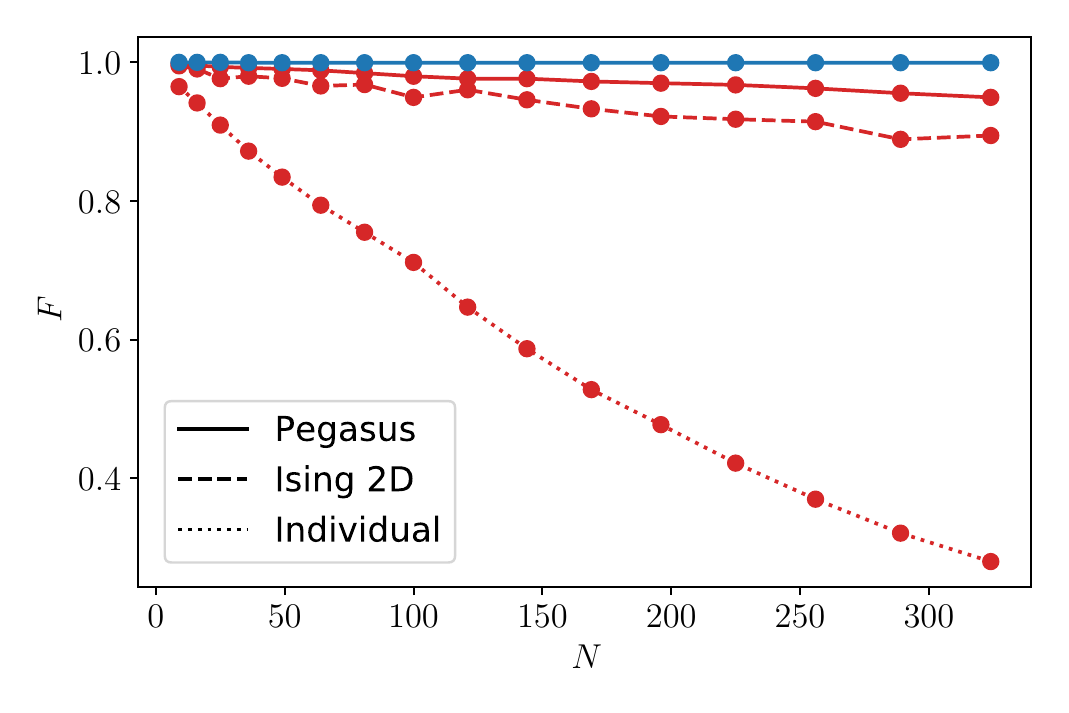}
    \caption{Measured global fidelity as a function of the number of qubits in the register. Red and blue curves refer, respectively to un-optimised and optimised CQIE protocols depicted in Fig. \ref{fig:driving}. Dotted line refers to individual reset ($\alpha_{i,j}=0$). Dashed line refers to the 2D square lattice with nearest neighbors interaction. Solid lines refer to maximal connectivity subgraphs available on the hardware.}
    \label{fig:fid}
\end{figure}
Figure \ref{fig:fid} reports the fidelity of preparation of the state $|00\dots 0\rangle $ as a function of the number $N$ of qubits in the quantum register, as measured for various protocols implemented onto a programmable network of superconducting qubits, for $N$ up to $N = 324$, and various adjacency matrices $\alpha_{i,j}$. Specifically, we implemented them on the D-Wave Advantage 6.4 processor with up to 5612 active qubits connected with the Pegasus topology \cite{pegasus}. The initial state of the network featured each qubit being randomly either in the up $|1\rangle $ or down state $|0\rangle $ of the Pauli operator $\sigma_z$ with equal probability. Each point is obtained by running the same protocol on the same network (as defined by its adjacency matrix $\alpha_{i,j}$), followed by the measurement of the Pauli operator $\sigma_z^i$ of each qubit, for a total of $\mathcal{N}=10^5$ times. The global fidelity was computed as the relative frequency of observing the target state $|0,0, \dots 0\rangle$. 

\begin{figure}
    \centering
    \includegraphics[width=\linewidth]{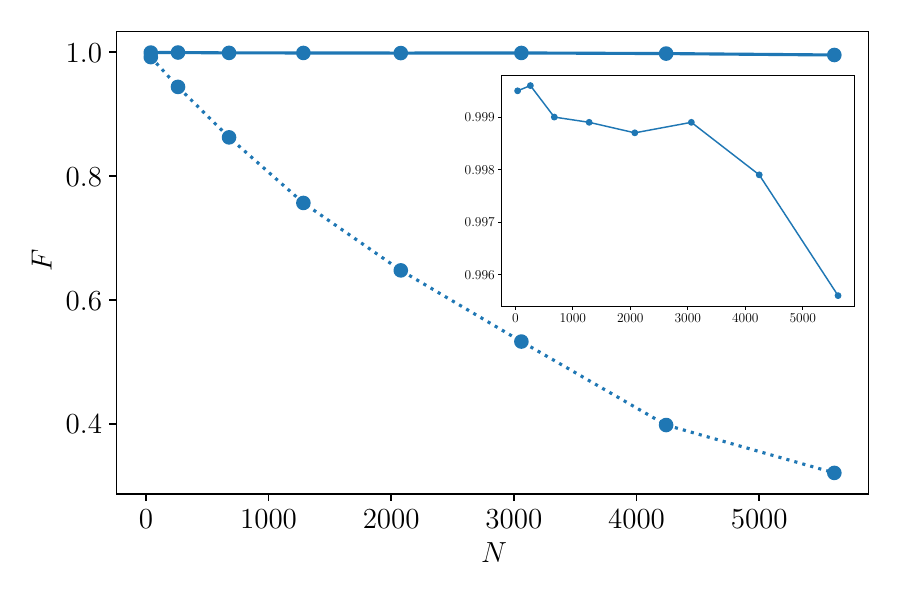}
    \caption{Measured global fidelity as a function of the number of qubits in the register. The solid blue curve is for the optimised cycle (see Fig. \ref{fig:driving}) run on maximal connectivity sub-graphs available on the hardware. The dotted blue curve is for the optimised individual reset cycle (i.e. featuring no interactions among the qubits $\alpha_{i,j}=0$. The inset shows a close-up of the solid curve. 
        }
    \label{fig:long_scaling}
\end{figure}

Both in Fig. \ref{fig:fid} and Fig. \ref{fig:long_scaling}, dotted curves refer to individual reset, i.e. $\alpha_{i,j}=0$ for all $i,j$'s;
Dashed curves refer to 2D nearest neighbour square lattice Ising connectivity; Solid curves refer to maximal connectivity subgraphs available on the hardware. Since the hardware connectivity is obtained by tiling together in the 2D plane of the chip patches of densely connected qubits, we used D-Wave's own graph generation tools \cite{dwdocs} to sample patches of increasing size from the Pegasus topology \cite{pegasus} and turned on all the available couplers of the chip inside these patches to achieve the maximal connectivity sub-networks of a given size. Such a network features larger connectivity (up to 15 neighbours per qubit)

Fig. \ref{fig:fid} evidences that the fidelity increases as the connectivity increases, as expected. With reference to to the three red curves (relative to the un-optimised protocol) note how fidelity is drastically increased when moving from the individual reset to the 2D nearest neighbour lattice. It further increases when moving to maximal connectivity. Finally, cycle optimisation further raises global fidelity which remains above the exceptional value of $0.9995$ for $N$ up to $324$ (blue solid curve).

Figure \ref{fig:long_scaling} shows the global fidelity as obtained by applying the optimised protocol onto maximal connectivity graphs of yet larger sizes. This resulted in exceptionally high global fidelities, with $F$ as large as $0.9956$ for $N=5612$, see blue solid line. For completeness, the network sizes were: $N=$40, 262, 678, 1284, 2078, 3061, 4241, 5612, and the according average connectivities were $c=$4.1, 6.02672, 6.56932, 6.81075, 6.93744, 7.01993, 7.08866, 7.14326, respectively.
For comparison, Fig. \ref{fig:long_scaling} reports as well the results obtained when applying the non-cooperative version of the optimised protocol (dotted blue line). Note how the fidelity dramatically drops down to about $0.35$ for the largest size $N$. We have checked that the data in fact obeys the expected Eq. (\ref{eq:F=f^N}), with $f$ being calculated from the data as the relative frequency over all runs and all qubits, to find any single qubit in the $|0\rangle$ state, which amounted to $f=1-(2.11 \pm 0.08) \times 10^{-4}$.

As a side remark, it is interesting to note that our optimisation not only improved the collective reset protocol but also resulted in an improved individual reset protocol. To see that compare the red dotted line in Fig.\ref{fig:fid} with the blue dotted line and Fig.\ref{fig:long_scaling}. 

\section{Conclusion and outlook}
By abandoning the customary method of preparing a quantum register via individual initialization of each qubit and embracing instead a global preparation approach, we have demonstrated here the achievement of exceptionally high values of global fidelities of the prepared state, that largely surpass the highest fidelities that can be currently obtained with individual preparation strategies. This was practically demonstrated on a programmable network of superconducting qubits, specifically a D-Wave Advantage 6.4 processor.

While this is an exceptional result \emph{per se}, its application to gate-based quantum computers is not immediate. While the presented method crucially rests on the existence of interactions with a thermal environment, gate-based quantum computers should be as isolated as possible from the environment. The present work then suggests that the next generations of quantum computers should be designed in such a way as to have tunable environment interactions so that they can efficiently perform both logically irreversible gates, e.g., reset to a specific state, which according to Landauer principle require dissipation \cite{Landauer61IBMRD5, buffoni2024generalized}, and logically reversible quantum gates, which require complete absence of environmental influences.
Meanwhile, since we are still in the so-called NISQ (noisy intermediate scale quantum) era \cite{Preskill18QUANTUM2} one can leverage on noise and attempt to implement the CQIE scheme on a current gate-based quantum computer, e.g., by means of Trotterisation techniques. Research in this direction is currently ongoing.

\section{Acknowledgments}
Access to D-Wave services was provided by CINECA
through the competitive ISCRA-C project ``COOPERA''.
L.B. was funded by PNRR MUR Project No. SOE0000098-ThermoQT financed by the European Union--Next Generation EU.

\bibliographystyle{unsrt}
\bibliography{bibliography}

\newpage
\newpage
\newpage

\newpage
\thispagestyle{empty}
\mbox{}
\newpage

\appendix
\section{The protocol} 
\begin{figure}
    \centering
    \includegraphics[width=0.9\linewidth]{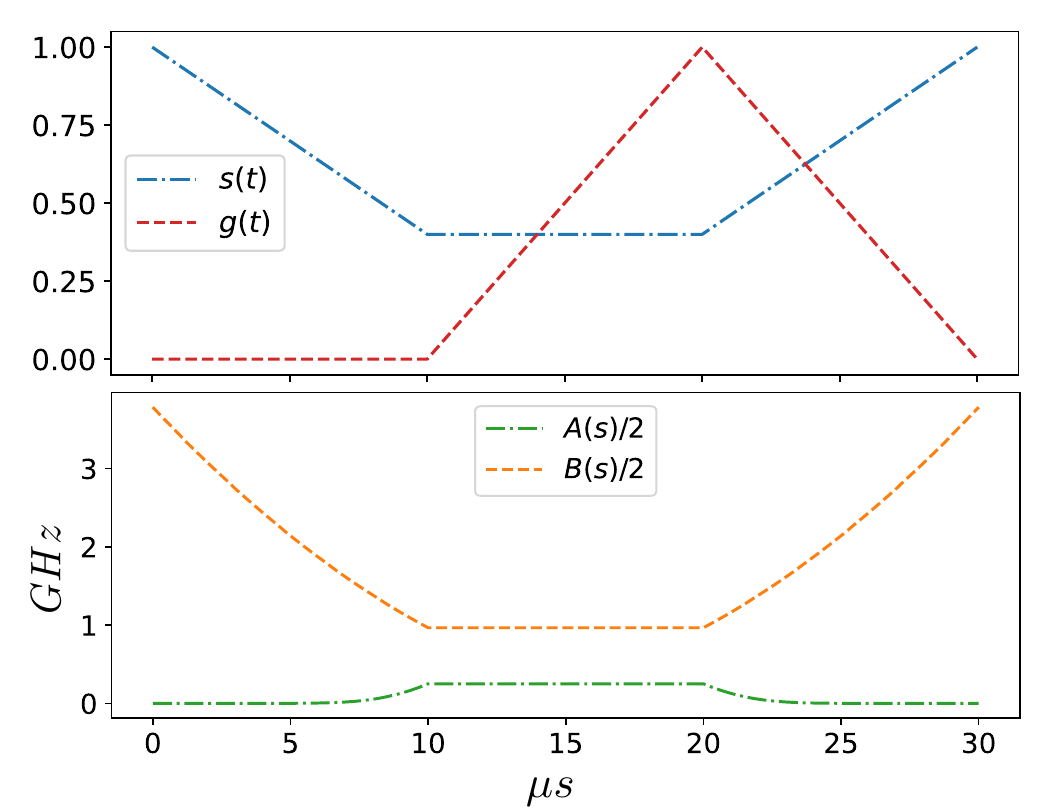}
    \caption{(Top panel) plot of the driving fields $s(t)$ and $g(t)$ during the protocol. (Bottom panel) the respective plot of the energies $A(s)/2$ and and $B(s)/2$ as a function of time. The quantum case is reported with the only difference, with respect to the classical case, that the term $A(s)/2$ is different from zero.}
    \label{fig:driving}
\end{figure}

In order to perform the collective erasure of \cite{buffoni2023cooperative} we need an Ising Hamiltonian that we can control with time-dependent fields across its critical point (see \cite{buffoni2023cooperative} for the details of the protocol). 
We implemented the protocol using the controllable spin Hamiltonian of the D-Wave Advantage quantum annealer that is:

\begin{align}
    \frac{H(t)}{h} = -\frac{A(s)}{2}\sum_i \sigma^x_i + \frac{B(s)}{2}\left [ g(t)\cdot \bar{h} \sum_i \sigma^z_i +  J \sum_{\langle i,j\rangle }  \sigma^z_i\sigma^z_j \right ],\\
    g(t)\; \text{and}\; s\equiv s(t) \in [0,1].
    \label{eq:advantage_ham}
\end{align}

where $s(t)$ and $g(t)$ are both piecewise linear functions of time and $A(s)$ and $B(s)$ are energies expressed in GHz units as given by the constructor (for reference see \cite{dwdocs,buffoni2023cooperative}). In particular to implement the desired time evolution we used the following protocol:
\begin{equation}
s(t) = \begin{cases} 
1-\frac{(1-\bar{s})t}{10}& 0 \mu s\leq t\leq 10 \mu s \\
\bar{s} & 10 \mu s\leq t\leq 20 \mu s \\
1-\frac{(1-\bar{s})(30-t)}{10} & 20 \mu s\leq t\leq 30 \mu s 
\end{cases}
\end{equation}
and
\begin{equation}
g(t) = \begin{cases} 
0 & 0 \mu s \leq t\leq 10 \mu s \\
\frac{t-10}{10} & 10 \mu s\leq t\leq 20 \mu s \\
\frac{30-t}{10} & 20 \mu s\leq t\leq 30 \mu s 
\end{cases}.
\end{equation}
That can be also seen more clearly pictured in Fig.\ref{fig:driving}.
In this description, it is worth noting that the transverse field component $A(s)$ becomes greater than zero, and thus the protocol becomes quantum, only for values for $\bar{s} \leq 0.5$. For the characterization below, we will take only two values $\bar{s} = 0.6$ (classical protocol) and $\bar{s} = 0.4$ (quantum protocol).
The other parameters that we will vary to characterize the protocol in the following are $\bar{h}$, a dimensionless value that determines the maximum local field applied to the qubits, and $J$, the dimensionless parameter determining the coupling between qubits. The connectivity $\langle i,j \rangle$ can be any subgraph of the full connectivity graph of the D-Wave Advantage 6.3 called Pegasus graph \cite{dwdocs} for which each qubit is connected on average to $15$ neighbors. To begin, we will start by considering square lattice graphs in order to leverage the well-known statistical properties of the 2D Ising model, but we will relax this assumption later in the paper and eventually take advantage of the full connectivity of the device.

In the main text, we used a modified version of $g(t)$ that is a quenched version  of the one employed above.
\begin{equation}\label{eq:g_quench}
    g(t)=
    \displaystyle{
    \begin{cases}
    0 & 0 \mu s\leq t\leq 10 \mu s \\
    \frac{t-10}{10} & 10 \mu s\leq t\leq 20 \mu s\\
    1 & 20 \mu s\leq t\leq 30 \mu s \\
    \frac{30.01 - t}{0.01} & 30 \mu s\leq t\leq 30.01 \mu s \,.
    \end{cases}
    }
\end{equation}

\section{Protocol characterization} To begin our characterization, we found the critical point of the 2D Ising model by sampling the equilibrium magnetization of the system at $\bar{h}=0$ and varying $J$. In this way, we located the second-order phase transition on the magnetization at $J=0.08 \pm 0.01$. This corresponds to an estimated temperature $T = 33 \pm 9$ mK using the well-known relation between critical temperature and coupling \cite{GoldenfeldBook}. An independent temperature estimate using the pseudo-likelihood method described in \cite{benedetti2016estimation} gave us a consistent estimate of $T= 31 \pm 4$ mK. 

\begin{figure*}
    \centering
    \includegraphics[width=\linewidth]{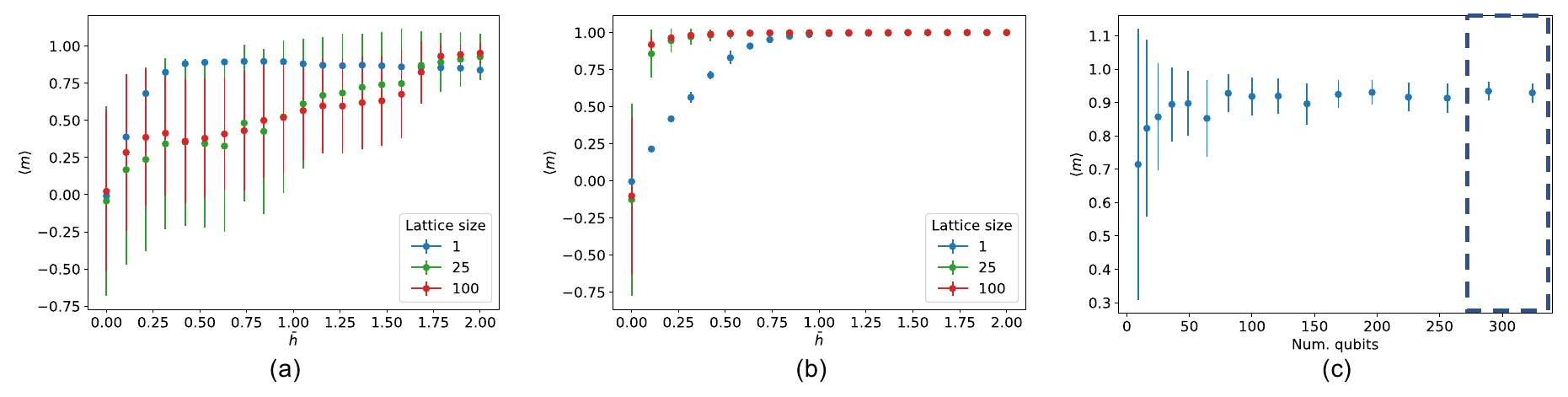}
    \caption{Panel (a): average magnetization $\langle m \rangle$ and corresponding standard deviation as a function of the local field strength $\bar{h}$ for the classical erasure protocol. A non-cooperative protocol (i.e. lattice size $1$) is compared against $5\times 5$ and $10 \times 10$ cooperative protocols on the Ising lattice. Panel (b): average magnetization and corresponding standard deviation as a function of the local field strength $\bar{h}$ for the quantum erasure protocol for various lattice sizes as explained above. Panel (c): average magnetization and corresponding standard deviationas a function of the number of qubits $N$ for a quantum cooperative protocol in an Ising lattice of increasing size (up to $18\times 18$). The dashed region marks the lattice size above which one has to resort to minor embedding techniques to fit the lattice into the fixed connectivity of the D-Wave Advantage 6.3 chip.}
    \label{fig:big_panel}
\end{figure*}

We then proceeded to a detailed investigation of the role that the various components play in the realization of this protocol. First, we fixed $J=0.12$ and varied the maximum longitudinal field $\bar{h}$ for both the classical and the quantum case and for lattices of various sizes spanning from the uncoupled (individual) reset of size $1$ up to a $10\times 10$ lattice. As a figure of merit, we used the average normalized magnetization of the system defined as $\langle m \rangle = \sum_i \sigma^z_i / N$. As can be seen in panels (a) and (b) of Fig.\ref{fig:big_panel} it is clear that the higher the longitudinal field employed the better the results. However, note that a larger field means more energy expenditure.
Regarding the cooperative effects, in the classical case, we do not see a big contribution from the cooperative effects, indeed the average magnetizations of the cooperative protocols are lower than the individual ones except for protocols in the high $\bar{h}$ regime. For the quantum case, we can nevertheless observe a clear effect on the quality of the result (i.e. higher average magnetization) given by cooperative effects as the size of the system increases. Note that the quantum protocol also magnetizes the system faster and with lower variance than the classical one. This is most probably because introducing the transverse field is very effective in decorrelating the spins and bringing them into a purely paramagnetic state, which makes it easy to align them in the same direction using the local field $\bar{h}$. On the contrary, the classical protocol cannot bring the coupling all the way to zero due to limitations on the control of $B(s)$. Due to the finite size of the system and hysteresis effects, the system will still have some magnetic domains and not be purely decorrelated and thus harder to align with the external field. 

To clarify this behaviour, we fixed the size of the system to a $10 \times 10$ lattice and $\bar{h}=0.25$ and varied the values of $J$ such that, during the protocol, the system was spending most of the time closer to the ferromagnetic phase or closer to the paramagnetic phase. The results of this experiment showed that the quantum protocol performs slightly better by increasing $J$ (i.e. being more ferromagnetic) while the classical protocol performs slightly better by decreasing $J$ (i.e. easier to decorrelate). This observation is coherent with the explanation given above, even if it is not a definitive proof for which we would need a device capable of implementing a classical protocol that can turn $B(s)$ to zero without introducing unwanted transverse fields.

To conclude the characterization of the protocol, we also investigated how, fixing all the other parameters, the quantum cooperative erasure scales with the size of the system. This observation is crucial for the following as if we want to use this protocol to reset multi-qubit registers the cooperative effects need to increase as the number of qubits increases. As one can see from panel (c) of Fig.\ref{fig:big_panel} a bigger system results in a higher final average magnetization after the protocol. Interestingly, we used this analysis to start relaxing the constraints on the 2D topology of the system. Indeed the last points of the graph, in the region contained by the dashed line, are not the result of a so-called minor embedding \cite{dwdocs} of the 2D Ising model on the physical connectivity of the chip. Indeed, it happens that above a certain size, one can no longer fit the 2D lattice onto the fixed Pegasus topology of the chip. The usual workaround for these situations is to perform a minor embedding by creating chains of qubits that are strongly ferromagnetically coupled and thus act as one unique logical qubit allowing one to circumvent the topological constraints of the chip. However, for our protocol, adding this minor embedding step means no longer working with an exact 2D Ising Hamiltonian with all its theoretical guarantees. Confirming that the protocol works just as well in the presence of minor embedding is crucial to move forward into the next section where we will then be able to move beyond the constraints of the 2D topology. 

\section{High fidelity registers} Having characterized all these properties of our protocol the main question we want to pose is whether this cooperative reset can be exploited to prepare high fidelity multi-qubit registers in an efficient way. We will then focus exclusively on the quantum protocol and taking a relatively high value of $\bar{h}=1$. 
The reset fidelity that we are interested in, is the overlap between the target reset state $\sigma=\ket{0}^{\otimes N}$ and the (unknown) state $\rho$ at the end of the protocol. Computing the fidelity then amounts to measuring the state of the system at the end of the protocol and counting the fraction of occurrences of the state $\ket{0}^{\otimes N}$. 

First, we established a baseline by computing the reset fidelity of the individual reset protocol, i.e. the protocol with $J=0$, for up to $N=324$ qubits. The results are shown as the dotted line in Fig.\ref{fig:fid}, where we can observe that the fidelity decreases quickly dropping below $0.5$ for a register of around $200$ qubits. Since we know that Eq.(\ref{eq:F=f^N}) can be interpreted as an effective temperature model as described in \cite{buffoni2022third} it is interesting to use the experimental fidelities to extract the effective temperature from the theoretical scaling: 

\begin{equation}
    F = \left ( 1 + e^{-\beta\Delta E} \right )^{-N}
    \label{eq:fidelity_scaling}.
\end{equation}

Indeed, we can fit the parameter beta above and obtain for the individual erasure an effective temperature of $T=32.56 \pm 0.06 mK$. This is consistent with our previous temperature estimates of the environment. 
We then implemented the full cooperative erasure protocol with $J=0.12$ for the Ising model and also with the full Pegasus topology to understand if having more connectivity would benefit the performance of the protocol with respect to the Ising case. The results can be seen as dashed and solid lines respectively in Fig.\ref{fig:fid}, where we see that in both cases we are able to prepare registers of dimensions up to $300$ qubits at fidelities $>90\%$ by exploiting the cooperative effects, in a regime where the individual erasure would have failed miserably.
We then fitted the effective temperature of Eq.(\ref{fig:fid}) on both cases. In contrast to the individual reset, the estimated temperature of the cooperative protocols on Ising and Pegasus are $T=23.1 \pm 0.1 mK$ and $T=20.78 \pm 0.05 mK$ respectively, which are significantly lower than the estimated initial temperature. In this regard, we can also see our protocol as an algorithmic cooling protocol whereby we have lowered the effective temperature of the ensemble of qubits.
In order to push the protocol even further we tried a variant of the protocol where the $g(t)$ is kept constant in the final stroke up until the last possible second and it gets quenched down to zero in $0.01 \mu s$ at the end, which is reported in Eq.\ref{eq:g_quench}. The aim is to make the system more robust against fluctuations. The result is the solid blue line of Fig.\ref{fig:fid} where we were able to obtain a $99.95 \%$ fidelity for the collective preparation of $324$ qubits corresponding to an effective temperature of $14.0 \pm 0.2 mK$ which is significantly below the measured temperature of ca 33 mK and even lower than D-Wave fridge nominal temperature that is reported to be $16.0 \pm 0.1 mK$ in its own datasheet \cite{dwdocs}.
It is interesting to notice that, while this protocol was originally implemented on a 2D Ising model to take advantage of the known phase transition, it works well also for the Pegasus topology with ferromagnetic interaction. It thus seems that we do not necessarily need to cross a phase transition for this cooperative effect to take place as the transverse field is good at decorrelating spins in itself without the need of entering a paramagnetic phase. We could then assume that, generally, the denser the topology, the better the outcome of the cooperative erasure.

\section{Scaling estimate}
\begin{figure}
    \centering
    \includegraphics[width=\linewidth]{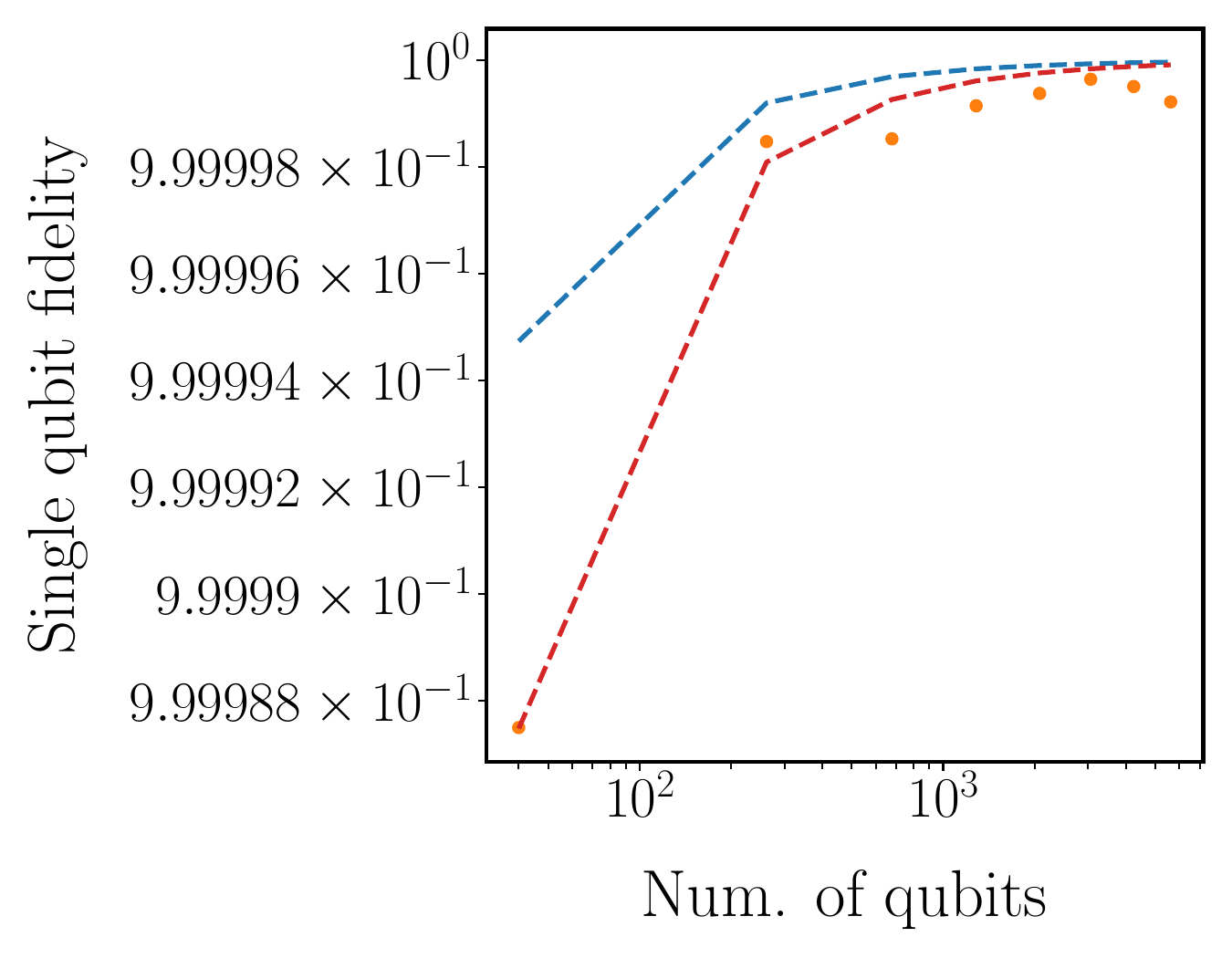}
    \caption{Average single-qubit fidelity computed measured for the qubits in the cooperative protocol (orange dots). The functions $(1-\alpha/N)$ (blue dashed line) and $(1-\hat{\alpha}/N)$ (red dashed line) are reported for visual reference regarding the scaling of $f(N)$ in the cooperative case.}
    \label{fig:single_fid}
\end{figure}

\begin{figure}
    \centering
    \includegraphics[width=\linewidth]{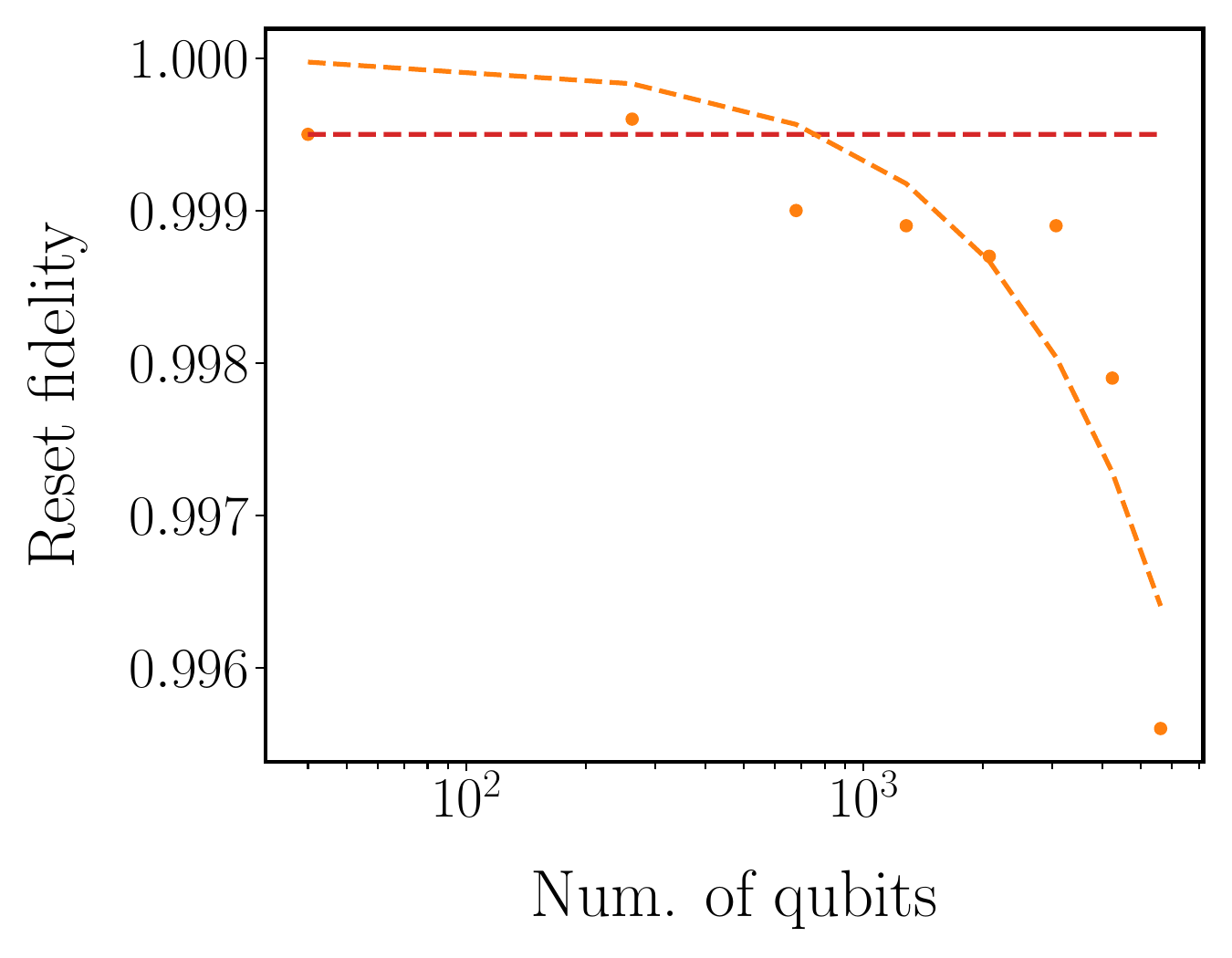}
    \caption{Measured global reset fidelity as a function of the number of qubits for the cooperative protocol (orange dots). The two possible models proposed above are reported as dashed lines $(1-\hat{\alpha}/N)^N$ (red) and $(1-\alpha)^{N/N_0}$ (orange).}
    \label{fig:just_coop}
\end{figure}

Knowing that for the individual (non-cooperative) case, the fidelity should scale as $F = f^N = (1-\alpha)^N$ we can fit the value of $\alpha$ from the blue data in Fig.\ref{fig:long_scaling} obtaining a value of $\alpha=2.11 \pm 0.08 \times 10^{-4}$. For the cooperative data, we can follow two approaches. The first comes by observing that in the cooperative case, we have that $f$ is not a constant but a function of the number of qubits (as can be observed from the orange points in Fig.\ref{fig:single_fid}). For example, we can assume that $f(N) = (1-\alpha/N)$, by using the same value of alpha found above we see that we should expect slightly higher single-qubit fidelities (blue dashed line in Fig.\ref{fig:single_fid}). But if we try to find the best fit of this functional form we get $f(N) = (1-\hat{\alpha}/N)$ with $\hat{\alpha}= 5.0 \pm 0.2 \times 10^{-4}$ (red dashed line in Fig.\ref{fig:single_fid}). However, what seems to be a good fit for the single qubit scaling turns out to be somehow unsatisfactory when dealing with the global fidelity. Indeed, as we can see in Fig.\ref{fig:just_coop} if we look at the measured values of the global fidelity (orange dots) the curve corresponding to the scaling $F=f(N)^N=(1-\hat{\alpha}/N)^N$ (red dashed line) doesn't look that good. We can then turn to the second approach, that is to consider $F=f^{N/N_0}=(1-\alpha)^{N/N_0}$ again we can fit the value of $N_0$ from the experimental data obtaining a value of $N_0=329 \pm 38$ (corresponding to the orange dashed line in Fig.\ref{fig:just_coop}) which looks like a better fit to the global fidelity.

\end{document}